\documentclass{article}
\usepackage{amssymb,amsmath,mathrsfs, amsfonts, graphicx,theorem,dsfont,yfonts, version, enumitem,stmaryrd}
\usepackage{authblk}
\usepackage{pst-all}
\usepackage{subcaption} 
\usepackage[english]{babel}
\usepackage{cancel}
\usepackage{bbold}
\usepackage{multirow}
\usepackage[normalem]{ulem}
\usepackage{hyperref}
\usepackage{cleveref}

\title{Clustering of temporal nodes profiles in dynamic networks of contacts}

\author[1,2]{Mehdi Djellabi}
\author[2]{Bertrand Jouve}

\affil[1]{IRIT (UMR 5505), Universit\'e Toulouse 1 Capitole, CNRS,118 Route de Narbonne, 31062 TOULOUSE Cedex 9, France.
djellabi.mehdi@gmail.com}
\affil[2]{LISST (UMR5193), Universit\'e de Toulouse, CNRS, Universit\'e Toulouse 2, 
5 all\'ees Antonio Machado, 31058 Toulouse Cedex 9, France. bertrand.jouve@univ-tlse2.fr}

\begin{document}
\maketitle

\begin{abstract}
{
Stream graphs are a very useful mode of representation for temporal network data, whose richness offers a wide range of possible approaches. The various methods aimed at generalising the classical approaches applied to static networks are constantly being improved. 
In this paper, we describe a framework that extend to stream graphs iterative weighted-rich-clubs characterisation for static networks proposed in \cite{dje20}.  The general principle is that we no longer consider the membership of a node to one of the weighted-rich-clubs for the whole time period, but each node is associated with a temporal profile which is the concatenation of the successive memberships of the node to the weighted-rich-clubs that appear, disappear and change all along the period. A clustering of these profiles gives the possibility to establish a reduced list of typical temporal profiles and so a more in-depth understanding of the temporal structure of the network.  This approach is tested on real world data produced by recording the interactions between different students within their respective schools.

\vspace{0.1cm}

\noindent \textbf{Temporal networks, Clustering, Dynamic patterns, Social network analysis }}
\end{abstract}
\newpage 
\section{Introduction}
Today, a large number of methods, measures and case studies have been published about structure analysis of static complex networks. Whatever these networks are directed, weighted or multi-layered, there are now efficient methods for modelling or mining their structure and a lot of them are implemented in usual softwares. At the beginning, in the 2000s, network science was mainly focused on the detection of macroscopic properties. It was shown, for example, that many real world networks had a ``small-world'' property \cite{watts1998} or that their degree distribution is a power-law \cite{barabasi1999}. However, these configurations can hide very different configurations \cite{amaral2000,broido2019} and complementary approaches quickly became necessary.

A certain number of studies have therefore focused on the search for particular configurations at lower levels, mesoscopic or microscopic, to explain macroscopic observations. 
For example the existence of different types of small world networks can be explained by a process that acts at the level of the node itself by limiting the number of links in the construction of the network \cite{amaral2000}. 
Or the community structure commonly found in social networks could explain the degree assortativity \cite{newman2003}. 
The search for communities in networks has been a very active area of research and, for a given network, the statistical features of its communities is a signature of the network's belonging to one of five types among information, communication, technological, biological, or social networks \cite{lancichinetti2010}. When the network hosts a dynamic process, and it is often the case for real world networks,  it makes sense to analyze network topology through the prism of the dynamics that can take place. 
Let us take a current topic, the spread of an epidemic. The existence of a community network structure influences the process of spreading an epidemic, but the internal structure of communities has little influence provided that they are dense enough in connections \cite{stegehuis2016}. Paradoxically, the way in which communities are articulated with each other has received relatively little attention from the community. 
If we want to find papers that study the way in which mesoscopic structures are organised within a network, we should look instead for Core-Periphery analyses, with properties as varied as rich-club, nested, or onion configuration \cite{csermely2013}, and in particular analyses that allow us to identify several cores \cite{rombach2014}. More recently \cite{dje20} have proposed an algorithm specifically dedicated to the search for sets of nodes that articulate communities.[Read, 2016] 

As mentioned above, the analysis of mesoscopic levels is particularly interesting for studying the dynamics that develop on a network. However, networks themselves are rarely static and very often contacts are made and broken with temporalities that can be very complex. It is therefore easy to understand that it is not possible, for example, to study the spread of an epidemic in a social network without taking into account the temporality of the links in the network. New challenges in network science therefore concern temporal networks or time-varying networks for which each edge is associated with one or more time intervals during which the links between the two nodes exist \cite{holme2019,masuda2020}. Conventional measurements used on static networks can no longer be used as is and must be adapted. At the mesoscopic level, the notion of community and community detection algorithms are defined as a function of time, with communities that change over time \cite{cazabet2019}. Several authors have also adapted the notions of Core-Periphery \cite{galimberti2021,sarkar2018,wu2015}.

In this paper we are concerned with the question of the organisation of a network at the mesoscopic level but in the context of temporal networks. We adapt \cite{dje20}'s algorithm to the case of temporal networks. By ranking the nodes according to a \emph{temporal strength} that takes into account both the degree of a node, some topological properties of the edges attached to it, and their change through time, we propose a temporal and multi-scale view of the network. From nodes which constitute in a way the core of the communities at any time to those which are very few connected to others, we highlight several types of nodes depending on their dynamical connectivity to the other.  
We use this approach to study two temporal social network datasets. The remainder of this paper is structured as follows: Section 2 is devoted to the notations and the method, Section 3 to a brief presentation of the datasets, Section 4 to the results and discussions of how our method provide a more advanced analysis of the data, and a conclusion as section 5.

\section{Method} 

\subsection{Notations}
A graph, or static graph, $G=(V,E)$ is a set $S$ of vertices (or nodes) associated with a set $E$ of unordered pairs of vertices. In this paper, graphs will be without loop or multiple edges. The \emph{open neighbourhood} of a vertex $v$ is $N(v)=\{u,(u,v) \in E\}$. For time varying graphs, we use the formalism and notations of a \emph{stream graph} \cite{latapy2018}. We denote by $S=(T,V,W,E)$ a (stream) graph where $T$ is a set of instants, $V$ a finite set of nodes, $W \subseteq T \times V$ a set of temporal nodes and $E= \subseteq T \times V \times V$ a set of temporal links. We suppose that all the graphs are not looped nor directed. We denote by $T_v=\{t,(t,v)\in W\}$ and $T_{uv}=\{t,(t,u,v)\in E\}$ the sets of time instants at which $v$ and $(u,v)$ are present respectively. 
In a symmetrical way, we denote by $V_t=\{v,(t,v)\in W\}$ and $E_t=\{(u,v),(t,u,v)\in E\}$ the sets of vertices and edges present at time $t$. At the instantaneous level, $G_t=(V_t,E_t)$ is  the graph induced by $S$ at time $t$. Of course, a link observed at time $t$ in $G_t$ may have begun before $t$ and may continue after $t$.
The \emph{instantaneous neighbourhood} $N_t$ of a vertex $v$ is defined by  $N_t(v)=\{(t,u),(t,(u,v)) \in E\}$ and is no more than the neighbourhood of $v$ in $G_t$.  
The \emph{instantaneous degree} $d_t(v)$ of a vertex $v$ is the cardinal $|N_t(v)|$ of $N_t(v)$. In the following, we will essentially use instantaneous neighbourhoods, but we will also refer to the neighbourhood of a vertex in a time windows $T'$ which is just $N_{T'}(v)=\bigcup_{T'}N_t(v)$. This is a disjoint union, and the cardinal of $N_{T'}(v)$ is just the sum of the cardinals of the $N_t(v)$. In particular, $N_T(v)$ is the neighbourhood of $v$ over the whole period $T$.   

Finally, notice that in the following we will often suppose that the set of nodes is fixed equal to $V$ and does not depend on time. So, in that case, we have $V=V_t$ for any $t\in T$.

\subsection{Temporal topological strength of a node} \label{section-topological}

Our aim is to adapt to a stream graph the \emph{topological strength} $\delta(v)$ defined in \cite{dje20} and efficient to explore the structure of links between densely connected parts of a network. This study is also a way to clarify, on a new example, how the measures built on static networks can be adapted to temporal networks. 
Remember that, on a static network $G=(V,E)$, the topological weight of an edge $(u,v)$ is define in \cite{dje20} as
\begin{equation}
 \omega(u,v) \approx \frac{|N(u)\cap N(v)| \cdot |N(u)| \cdot |N(v)|}{|N(u)| + |N(v)|} \;\;\;\; \text{if} \;\;(u,v) \in E \;\; \text{and} \;\; 0 \;\; \text{otherwise}
 \label{omega}
 \end{equation}
 and the topological strength of a vertex $i$ is
\begin{equation} 
\delta(u)=\sum_{v\in N(u)}\omega(u,v)
 \label{delta}
 \end{equation}
where $N(i)$ is the open neighbourhood of $i$ in $G$ and $\;\approx \;$ signifies an equality up to a multiplicative scalar factor. The topological strength is a measure of relative density of links in the neighbourhood of a vertex taking into account both the degree of the vertex and of its neighbours and the ratio of common neighbours between
the vertex and its neighbours. The iterative search of $\delta$-weighted rich-clubs [ref] provide several hierarchical layers of nodes, which altogether constitute
what is called the `dense part' of the network. The set of vertices that are not in any of the layers is called the `sparse part' (see \cite{dje20} for details on the algorithm).

Basically, there are two ways to adapt such a measure to a stream graph $S=(T,V,W,E)$. The first option is to calculate one value for each pair $(u,v)$ for the whole period $T$ of the graph. All you have to do is to replace $N$ and $E$ by $N_T$  and $E_T$ in equation \ref{omega}. In that case, the purpose is in some way rather to obtain a mean value of $\omega$ over the whole period $T$ for each edge. Even if the sets of neighbours and their intersections are time dependent, two edges can be ``active'' (ie. with a high weight) at different time periods and have the same weight value. The same is true for vertices and $\delta$. Therefore, this solution may be efficient to classify the vertices depending on their level of strength over the whole period of study, but can never be efficient to differentiate two vertices with similar strengths but ``active'' at different instants.       

In this paper we adopt a second solution that leads to a profile of \emph{instantaneous topological strength} for each vertex over the period. The instantaneous topological strength $\delta_t$ is defined as the topological strength on $G_t$. It is obtained from equations \ref{omega} and \ref{delta} by replacing $N$ by $N_t$: 
\begin{equation}
    \omega_t(u,v) \approx \frac{|N_t(u)\cap N_t(v)| \cdot |N_t(u)| \cdot |N_t(v)|}{|N_t(u)| + |N_t(v)|} \;\;\;\; \text{if} \;\;(u,v) \in E_t \;\; \text{and} \;\; 0 \;\; \text{otherwise}
\label{eq:omega_instant1}
\end{equation}
and 
\begin{equation} 
\delta_t(u)=\sum_{v\in N_t(u)}\omega_t(u,v)
 \label{delta_instant}
 \end{equation}

At a given instant $t$, for real world networks, the graphs $G_t$ are often sparse in connections and then may display very different topologies with few continuity. The result are temporal series of graphs $G_t$ and of values of $\omega_t$ and $\delta_t$ difficult to analyse. A way to overcome this difficulty is to no longer consider isolated instants $t$ in $T$ but a series of $n$ successive time windows $T_{i}=[t_i,t_{i+1}]$ of a small duration $t_{i+1}-t_i=\Delta$ that cover $T$, and to take the mean of $\omega_t$ over each time window $T_i$. 
The result is two smoothing values $\overline{\omega}_{T_i}$ and $\overline{\delta}_{T_i}$ associated to each time window $T_{i}$:

\begin{equation}
    \overline{\omega}_{T_{i}}(u,v)  =  \frac{1}{\Delta}  \int_{T_{i}} \omega_t(u,v) \,dt \;\;\;\;\;\; \text{and} \;\;\;\;\;\; \overline{\delta}_{T_{i}}(u)= \displaystyle \sum_{\substack{u\in N_t(v) \\ t\in T_{i}}}\overline{\omega}_{T_{i}}(u,v)
    \label{eq:simple_omega_continu}
\end{equation}
 
For each node,  $(\overline{\delta}_{T_{i}})_{1\leq i \leq n}$ is a profile of activity that reflects the dynamics of each vertex with its topological neighbourhood all over the period $T$.

\subsection{A similarity measure between temporal profiles}

To find the $\delta$-weighted-rich-clubs, we use the \emph{ItRich} algorithm described in \cite{dje20} with $\overline{\delta}_{T_{i}}$ for each of the $T_i$. Remind that this algorithm is an iterative process that provides a series of rich-clubs for $\delta$, that is sets of nodes with the highest $\delta$ and whose sum of the weights $\omega$ of the edges linking them is high comparing to a well chosen null model. 
The union of these rich-clubs is called the dense part of the network and the rest is the sparse part. 
Whereas in static graphs isolated nodes can easily be ignored when the focus is on links configuration, with stream graphs a node can be isolated at some instants and have links at some others. So, such nodes cannot be kept out and they will be classified as \emph{passive nodes} at each instant for which they have a zero instantaneous degree. 
To sum up, at each instant $t$ the set of nodes of the stream graph can be partitioned into 3 subsets: the subsets $D_t$ and $S_t$ of nodes that belong to the dense and sparse parts respectively (calculated for the time-window $T_i$ containing the instant $t$), and the set $P_t$ of passive nodes. Thus, the set of temporal nodes $W$ of a stream graph is partitioned into 3 subsets
$$D=\bigcup_{t\in T} D_t \hspace{.6cm}, \hspace{.6cm} S=\bigcup_{t\in T} S_t \hspace{.6cm}, \hspace{.6cm} P=\bigcup_{t\in T} P_t.$$

In the following we will consider that $T$ is a countable set, which is often a good approximation with real world data whose continuity is anyway only reconstructed from discontinuous measurements. But, the method is easily adaptable to computer algebra. 
So, each node $u$ is associated with a profile which is a word (possibly of infinite length) with a 3-letter alphabet $\{D,S,P\}$, each letter corresponding to the classification of the node as dense, non-dense or passive at each instant. 
Unlike the static case, this method does not give a classification of the nodes but a series of temporal profiles to be classified according to their similarity. 

Set $R^{(u)} = \{r_1^{(u)},r_2^{(u)},...,\cdots \}$ the temporal profile of a node $u$ with $r_t^{(u)} \in \{D_t,S_t,P_t\}$ for all $t \in T$, a large number of methods is available to compare two words, largely depending on their representation. To code our 3-letter words, we have chosen to use a multidimensional binary indicator vector, efficient in genome studies \cite{yin2015improved}. In our case, each profile $R^{(u)}$ is converted into three binary vectors $R^{(u)}_{D}$, $R^{(u)}_{S}$ and $R^{(u)}_{P}$, with 
\begin{equation}
\forall t\in T, \;\; R^{(u)}_{D}(t) = \left\{
    \begin{array}{ll}
        1 & \mbox{if } r_t^{(u)} = D_t \\
        0 & \mbox{otherwise}
    \end{array}
\right.
\label{eq:signal}
\end{equation}
and $R^{(u)}_{S}$ and $R^{(u)}_{P}$ obtained in the same way.

Then, given a similarity measure $s$ between binary vectors, we define the similarity $\overline{s}$ between $R^{(u)}$ and $R^{(v)}$ as a weighted average of the similarities between the three vectors: 
\begin{equation}
\overline{s}(R^{(u)}, R^{(v)}) = \alpha_D \cdot s(R^{(u)}_{D},R^{(v)}_{D})+ \alpha_S \cdot s(R^{(u)}_{S},R^{(v)}_{S})+\alpha_P \cdot s(R^{(u)}_{P},R^{(v)}_{P})
\label{eq:sim}
\end{equation}
with $ \alpha_D+ \alpha_S+ \alpha_P =1$. 
For the applications below, we chose the simple case where $ \alpha_D= \alpha_S= \alpha_P =\frac{1}{3}$, but another choice may be more suitable depending on the data. 
About the similarity $s$ between the different binary vectors, we will chose the cosine similarity $s(A,B) = \frac{<A,B>}{||A||\cdot ||B||}$ where $<\cdot,\cdot>$ is an inner product and $||\cdot||$ the associated norm.

\section{Data sets} 
In this section we present the data on which we apply the method described above. It consists of three datasets that can be found on the site \url{http://www.sociopatterns.org}, which are the results of recordings the interactions between pupils within their respective schools. Each individual is equipped with a badge containing an RFID (Radio Frequency Identification) type chip, which registers a contact each time it is close to another chip. The students are required to wear their badges on their chests, so that interactions are only recorded when two people are in front of each other, and at a maximum range of $1m$ to $1.5m$. This last restriction was added in order to be able to analyse short range interactions. The infrastructure settings are adjusted so that the proximity between two people wearing the RFID badges can be assessed and recorded with a probability of more than $99\%$. 
This time scale allows an adequate description of interactions between individuals, even if they are brief. The interactions that occur outside the school premises are not recorded.

 We have analysed two of these three datasets. The first one is a series of recordings made in a primary school with 232 pupils and 10 teachers spread over 10 classrooms, spread over 5 different school levels. The number of pupils in each class varies between 21 and 26 in total. The second dataset is a series of recordings from students in preparatory classes \cite{barrat14} and which were recorded over a whole day. The number of students in this last dataset is 327, and they are spread over 9 different classrooms with a number of students ranging from 29 to 40. As data are collected using the same device, it is possible to compare their results. For both datasets, we choose the day with the most variability so that our approach is emphasised. It is the first day for the first dataset and the third one for the second dataset.  
 
 Recordings are made every $\Delta_t=20$ seconds, so that resolution limit of the data makes it possible to generate dynamic graphs with a minimum window width of 20 seconds. 
 But as explained before, it is efficient to smooth these relations over a larger period. 
 We choose to aggregate the data in 5 minutes time-windows, which corresponds to $p=15$ time steps of 20 seconds and to $\Delta=5$ minutes in \cref{eq:simple_omega_continu}. 
 This choice is guided by the average duration of the moments of the day during which a large number of interactions are recorded. Indeed, during breaks in recess, the number of interactions increases considerably, which motivates the choice of a time window that should not be wider than the duration of the break. And as these breaks last about 15 minutes (preparatory classes) and 25 minutes (primary school), it is important to take a time window that allows for several graphs to be obtained during these breaks. On the other hand, the time window should not be too short either, as pupils are in their classrooms for most of the day, and interactions are more rarely recorded there. Thus, a window that is too short would result in a series of graphs with too few links allowing us to derive usable information later on. 

From an implementation point of view, the resolution limit means that $\overline{\omega}_{T_{i}}$ is constant over any 20s time-windows, and so we have for each time-window $T_i$:

\begin{equation}
    \overline{\omega}_{T_i}(u,v) 
    =  \frac{1}{p}  \sum_{k=0}^{p-1} \omega_{t_i+k\Delta_t}(u,v)
    \label{eq:simple_omega_continu_1}
\end{equation}
with $\Delta=p\Delta_t$. For one-day primary school and preparatory classes data, each profile is a $\{D,S,P\}$-word of length 1530 and 1620 respectively.

\section{Results and discussion} 

\subsection{Description of the results}

\begin{figure}
  \begin{subfigure}[b]{0.49\textwidth}
    \resizebox{1.\textwidth}{.31\textheight}{\includegraphics{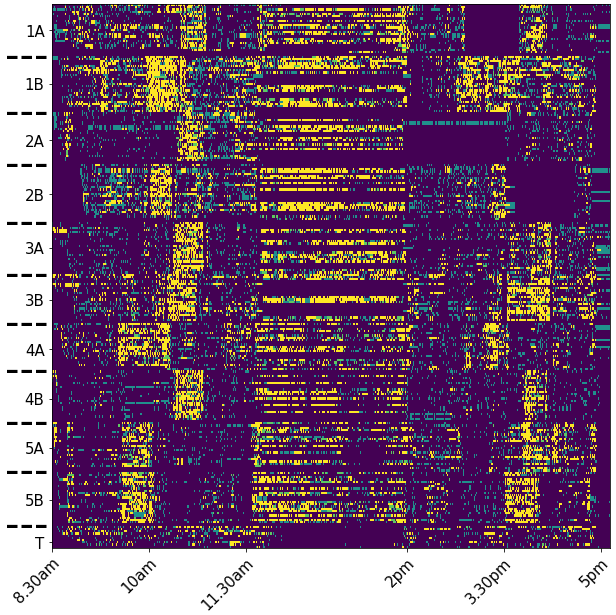}}
    \caption{}
    \label{day-prim}
  \end{subfigure}
  \begin{subfigure}[b]{0.49\textwidth}
    \resizebox{1.\textwidth}{.31\textheight}{\includegraphics{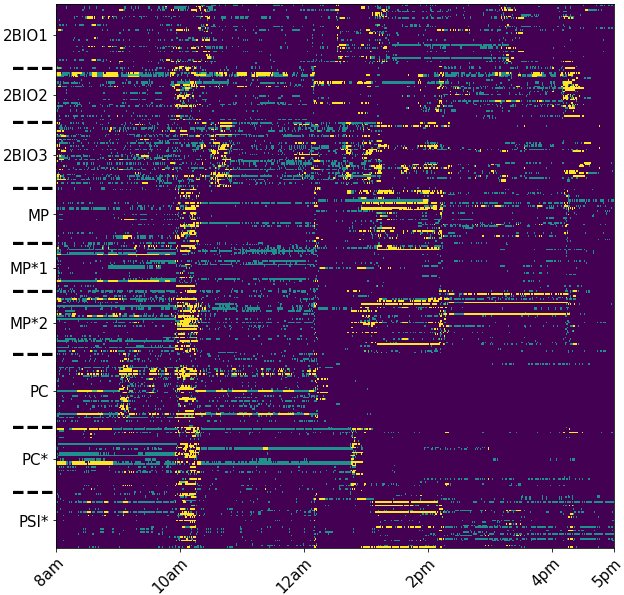}}
    \caption{}
    \label{day-lyce}
  \end{subfigure} 
\caption{One day $(D,S,P)$-dynamics of primary school (a) and preparatory classes (b) students. Each row is associated with a student and rows are ordered by class (indicated by the left scale in each graphics). The yellow colour indicates the belonging to the dense part and the green to the sparse part. The purple colour indicates the passive state of the concerned individual within the corresponding $\Delta_t$-time-window.}
\label{fig:day}
\end{figure}

The dynamics displayed by the pupils in the primary school and in the preparatory classes are very different (\cref{fig:day}).  It is clear that the average rate of belonging to the dense part is higher for the primary school pupils, with a value equal to $12\%$ for the whole day against $3\%$ for the preparatory classes pupils. Note that the average rate of belonging to the sparse part of each pupil is almost the same in both sets of data, and represents respectively $9.05\%$ and $9.2\%$ of the total length of the day. The passivity rate is more important among preparatory school students, and represents $87.53\%$ of the day compared to $77\%$ for primary school pupils.
\newline This can be explained by the fact that, unlike primary school pupils, students in preparatory classes are in a more demanding school context which leaves them less time to form groups and discuss with each other for long periods of time, which induces less density in the temporal network. 
\newline Another aspect that differentiates the two dynamics is the fact that the recreational breaks take place at the same time for the majority of the classes in the preparatory school (with the exception of the $2BIO3$ class, whose break took place half an hour later than those of the other classes), which is not the case for the breaks in the primary school, that are staggered from one class to another. 

In order to check the extent to which the time that a given student $u$ spends in the dense part or in the sparse part is directly related to the number of her interactions, we plot, for each data set, the average degree of $u$ against the following membership rates 

\begin{equation}
    \tau_A(u) = 1-\frac{\Theta_P(u)}{|T|} \hspace{1,5cm}
    \tau_D(u) = \frac{\Theta_D(u)}{|T|} \hspace{1,5cm}
    \tau_S(u) = \frac{\Theta_S(u)}{|T|}
%
\end{equation}
where $\Theta_D(u),\Theta_S(u)$ and $\Theta_P(u)$ are the number of instants $t$ for which node $u$ is classified in $D$, $S$, and $P$ respectively. The average degree of a node $u$ over $T$ is
\begin{equation}
\overline{d}(u)=\frac{1}{|T|} \int d_t(u)du 
\end{equation}
Please note that $\tau_A = \tau_D +  \tau_S$ for all node $u$, which means that $\tau_A$ is a global measure of interaction duration of a node whatever it is in the dense or in the sparse part.    

\begin{figure}
\begin{centering}
  \begin{subfigure}[b]{0.4\textwidth}
    \includegraphics[width=\textwidth]{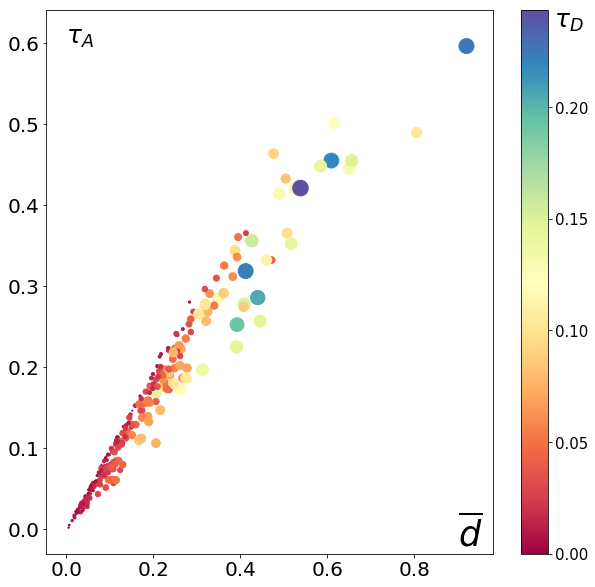}
    \caption{}
    \label{nuage-a}
  \end{subfigure}
  \begin{subfigure}[b]{0.4\textwidth}
   \includegraphics[width=\textwidth]{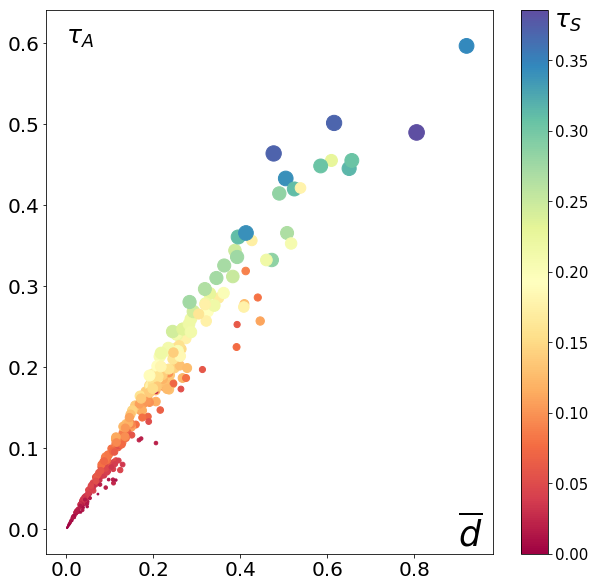}
    \caption{}
    \label{nuage-b}
  \end{subfigure}

  \begin{subfigure}[b]{0.4\textwidth}
   \includegraphics[width=\textwidth]{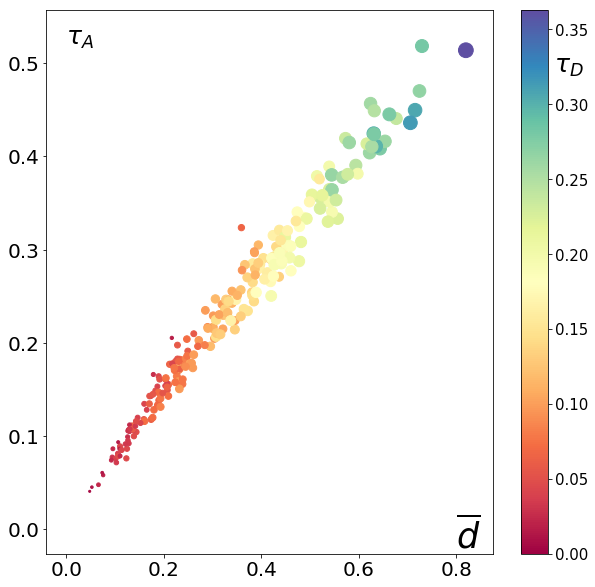}
    \caption{}
    \label{nuage-c}
  \end{subfigure} 
  \begin{subfigure}[b]{0.4\textwidth}
    \includegraphics[width=\textwidth]{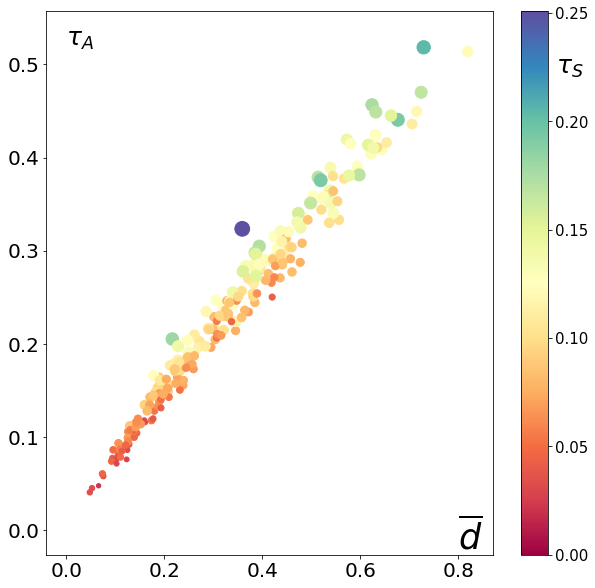}
    \caption{}
    \label{nuage-d}
  \end{subfigure}
\caption{(a) and (b) displays the heat map of $\tau_D$ and $\tau_S$ on the scatter-plot representing $\tau_A$ \textit{vs.}
$\overline{d}$ for preparatory class students data. The same diagrams are reproduced respectively in (c) and (d) for the data of primary school students.}
\label{fig:nuage}
\end{centering}
\end{figure}

We observe on the \cref{fig:nuage} that for both datasets, there is a strong correlation between the interaction duration $\tau_A$ of the nodes and their average degree $\overline{d}$ (the Pearson's correlation coefficient is 0.98 in the data set of the preparatory classes and 0.99 in the primary school data set). However, this correlation is, for example, less important between $\tau_D$ and $\overline{d}$ with a coefficient that drops to 0.81 in the case of the preparatory classes data set. In fact, we can observe on \cref{nuage-a} that for this latter data set, the student with the highest membership rate in the dense part $\tau_D$ has an average degree $\overline{d}$ of $0.53$, which is clearly less than the maximum value observed, which reaches $0.91$. This effect is not observed on the primary school data set for $\tau_D$, but it can be seen on the \cref{nuage-d} for the $\tau_S$ values, where the pupil with the highest membership rate in the sparse part is neither the one with the highest average degree nor the smallest. 
\newline  This can be explained by the fact that, applied on short time windows (5mn), the ItRich algorithm is able to reveal students who interact more than average during moments of high densities (recess, lunch breaks) and less during class hours. Conversely, there are students who form links even during class hours, and given the low density of interactions at these times, their average degree increases while their membership rate to the dense part remains low. This would suggest that there are several complex dynamic topological patterns within the studied datasets, that we propose to highlight in the following section.
 
\subsection{Clustering of similar profiles}
\begin{figure}
\begin{centering}
  \begin{subfigure}[b]{0.4\textwidth}
    \includegraphics[width=\textwidth]{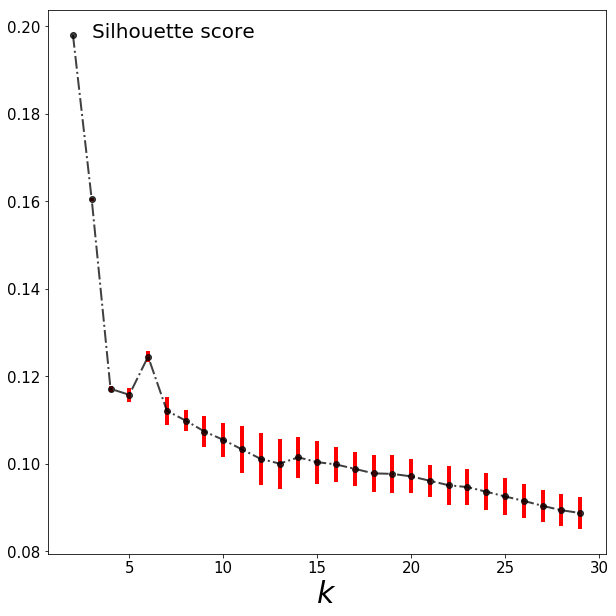}
    \caption{primary school data}
    \label{silhouette-a}
  \end{subfigure}
  \begin{subfigure}[b]{0.4\textwidth}
   \includegraphics[width=\textwidth]{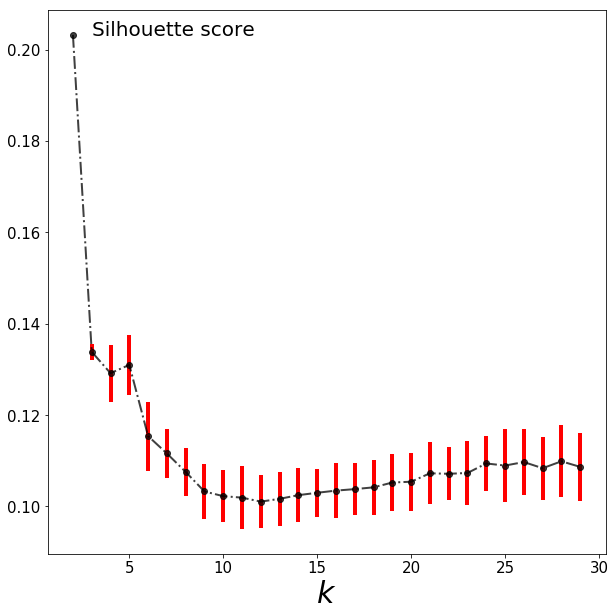}
    \caption{highschool data}
    \label{silhouette-b}
  \end{subfigure}

\caption{The average silhouette score as a function of the number of clusters $k$ for the two datasets under study, with 100 runs of the k-means algorithm for each value of $k$.}
\label{fig:silhouette}
\end{centering}
\end{figure}

We compute the similarity matrix between 3-letter profiles using the expression given by \cref{eq:sim} with a coefficient of one third for each term (none of the similarities is preferred to the two others) and the dot product as inner product. For clustering, we use the
k-means algorithm and choose the number of clusters with the silhouette method \cite{ROUSSEEUW198753}. This method calculates an \emph{average silhouette score} $\overline{s}(k) \in [-1;1]$ that estimates the quality of a partition into $k$ clusters through the difference between the mean intra-cluster distance $a(u)$ of a point $u$ (the average distance from $u$ to the points of the cluster it has been assigned to) and the mean nearest-cluster distance $b(u)$ (the average distance to the nearest cluster to which the point $u$ does not belong). 
For each data set and $k$ clusters, the average silhouette score is:

$$ \overline{s}(k)=\frac1{|V|} \sum_{u \in V}\;\frac{b(u)-a(u)}{\max\{a(u),b(u)\}} $$
We calculate this coefficient for several values of $k$ and select a value of $k$ that has the highest average silhouette score. Because the k-means algorithm depends on an initially arbitrary order which can modify its output, $100$ values of $\overline{s}(k)$ are averaged for each value of $k$. 
The results are shown in the \cref{fig:silhouette}.

\begin{figure}
  \begin{subfigure}[b]{0.49\textwidth}
    \resizebox{1.\textwidth}{.31\textheight}{\includegraphics{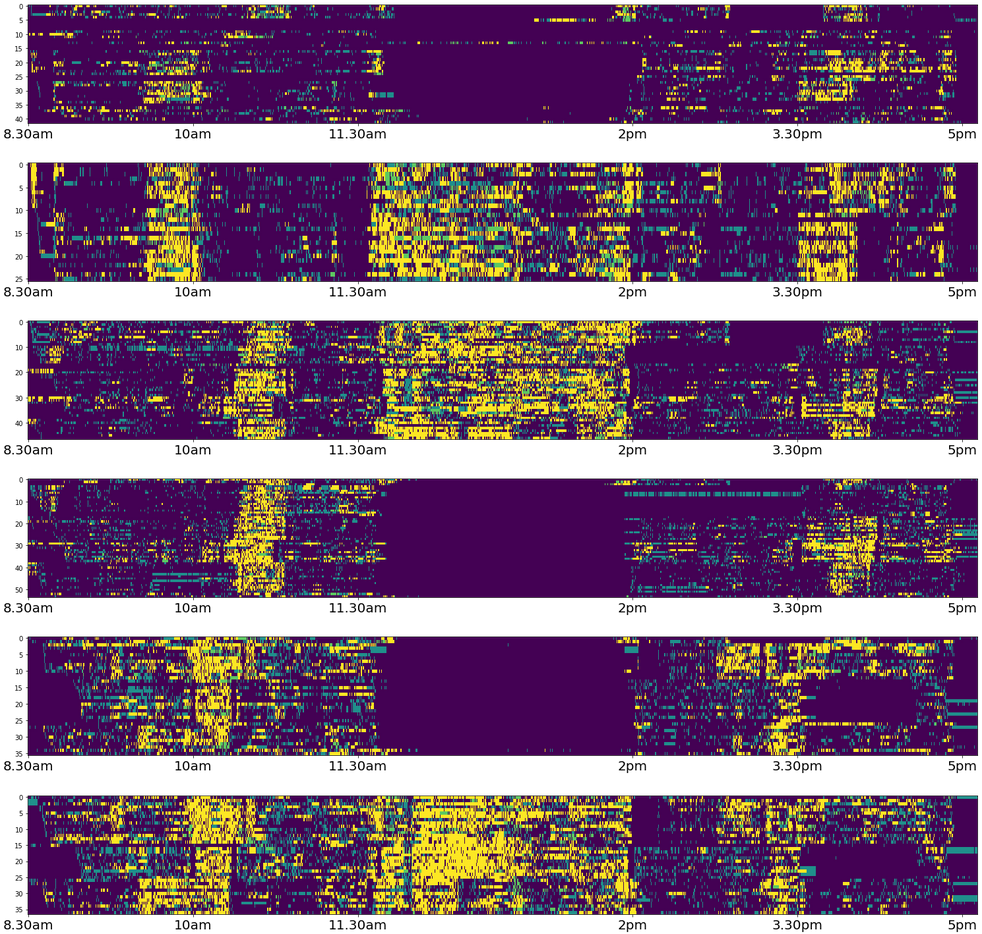}}
    \caption{Primary school data}
    \label{clusters-prim}
  \end{subfigure}
  \begin{subfigure}[b]{0.49\textwidth}
    \resizebox{1.\textwidth}{.31\textheight}{\includegraphics{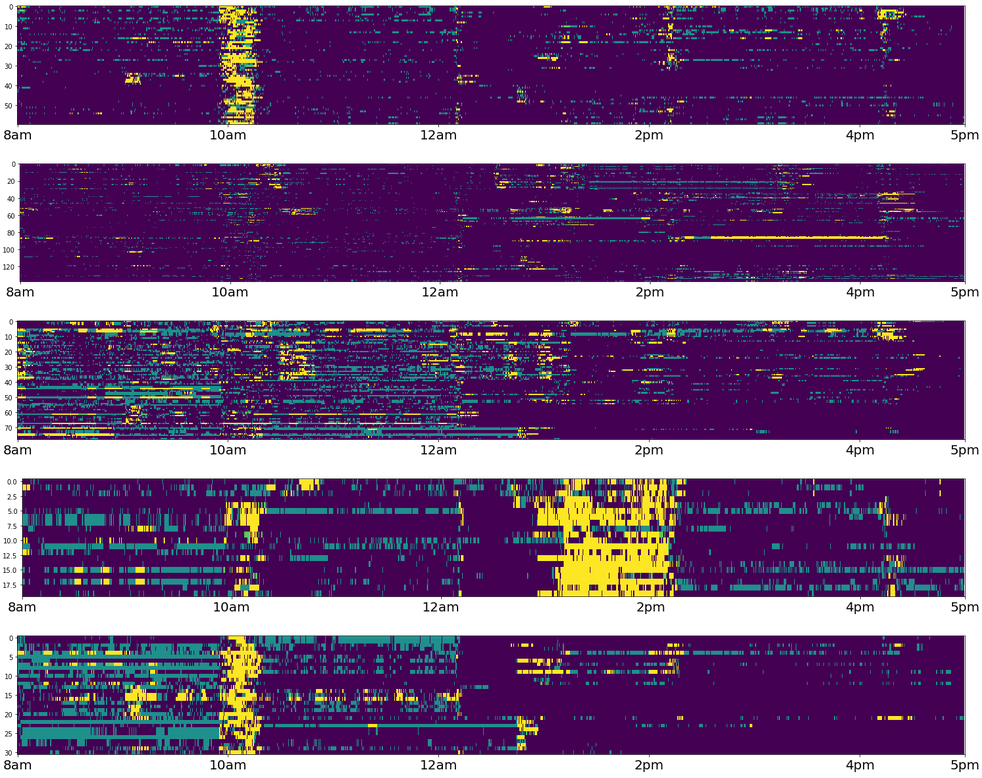}}
    \caption{Highschool data }
    \label{clusters-lycee}
  \end{subfigure} 
\caption{Classification of $(D,S,P)$-temporal-profils of the students with a k-means method (6 clusters for $(a)$ and 5 for $(b)$).}
\label{fig:clusters}
\end{figure}

We observe that the maximum score is obtained for low values of $k$ (number of clusters equal to two). However, the two corresponding clusters are not interesting to interpret because they contain respectively the individuals who showed high activity during the day, and those who interacted weakly (those who were in a passive state for the major part of the day). It is therefore more interesting to choose a value of $k$ which is a good compromise between the variety of the results (diversity of the extracted profiles) and their accuracy. For this, we have chosen a number of clusters equal to $6$ for the primary school data set and $5$ for the preparatory classes data set, each one corresponding to a local maximum of the curves. The students $\delta$-profiles grouped according to the different clusters for these values of $k$ are displayed on the \cref{fig:clusters}, and the distribution of the students of the different groups within each class is given in \cref{table:table}.

\begin{table}[!ht]
\resizebox{\textwidth}{1.2cm}{
\begin{tabular}{c|c|c|c|c|c|c|c|c|c|c|c|cccccccccc}
\cline{2-21}
                                  & \multicolumn{11}{c|}{\textbf{Primary school data set}}                                                                                                 & \multicolumn{9}{c|}{\textbf{Highschool data set}}                                                             \\ \cline{2-21} 
                                  & \textbf{1A} & \textbf{1B} & \textbf{2A} & \textbf{2B} & \textbf{3A} & \textbf{3B} & \textbf{4A} & \textbf{4B} & \textbf{5A} & \textbf{5B} & \textbf{T} & \multicolumn{1}{c|}{\textbf{2B1}} & \multicolumn{1}{c|}{\textbf{2B2}} & \multicolumn{1}{c|}{\textbf{2B3}} & \multicolumn{1}{c|}{\textbf{MP}} & \multicolumn{1}{c|}{\textbf{MP*1}} & \multicolumn{1}{c|}{\textbf{MP*2}} & \multicolumn{1}{c|}{\textbf{PC}} & \multicolumn{1}{c|}{\textbf{PC*}} & \multicolumn{1}{c|}{\textbf{PSI*}} \\ \hline
\multicolumn{1}{|c|}{\textbf{C1}} & 7           & 0           & 1           & 1           & 1           & 3           & 2           & 1           & 11          & 9           & 6          & \multicolumn{1}{c|}{1}              & \multicolumn{1}{c|}{7}              & \multicolumn{1}{c|}{0}              & \multicolumn{1}{c|}{10}          & \multicolumn{1}{c|}{4}             & \multicolumn{1}{c|}{11}            & \multicolumn{1}{c|}{7}           & \multicolumn{1}{c|}{9}            & \multicolumn{1}{c|}{11}  & \multicolumn{1}{c|}{\textbf{C'1}}            \\ \hline
\multicolumn{1}{|c|}{\textbf{C2}} & 0           & 0           & 0           & 0           & 0           & 0           & 0           & 0           & 11          & 15          & 0          & \multicolumn{1}{c|}{32}             & \multicolumn{1}{c|}{16}             & \multicolumn{1}{c|}{12}             & \multicolumn{1}{c|}{14}          & \multicolumn{1}{c|}{11}            & \multicolumn{1}{c|}{8}             & \multicolumn{1}{c|}{15}          & \multicolumn{1}{c|}{14}           & \multicolumn{1}{c|}{16}      & \multicolumn{1}{c|}{\textbf{C'2}}      \\ \hline
\multicolumn{1}{|c|}{\textbf{C3}} & 10          & 0           & 7           & 2           & 11          & 8           & 1           & 8           & 0           & 0           & 0          & \multicolumn{1}{c|}{4}              & \multicolumn{1}{c|}{10}             & \multicolumn{1}{c|}{25}             & \multicolumn{1}{c|}{2}           & \multicolumn{1}{c|}{10}            & \multicolumn{1}{c|}{5}             & \multicolumn{1}{c|}{14}          & \multicolumn{1}{c|}{7}            & \multicolumn{1}{c|}{1}        & \multicolumn{1}{c|}{\textbf{C'3}}     \\ \hline
\multicolumn{1}{|c|}{\textbf{C4}} & 3           & 0           & 14          & 0           & 11          & 10          & 0           & 14          & 0           & 0           & 2          & \multicolumn{1}{c|}{0}              & \multicolumn{1}{c|}{0}              & \multicolumn{1}{c|}{3}              & \multicolumn{1}{c|}{7}           & \multicolumn{1}{c|}{2}             & \multicolumn{1}{c|}{3}             & \multicolumn{1}{c|}{0}           & \multicolumn{1}{c|}{0}            & \multicolumn{1}{c|}{5}         & \multicolumn{1}{c|}{\textbf{C'4}}    \\ \hline
\multicolumn{1}{|c|}{\textbf{C5}} & 2           & 11          & 0           & 13          & 0           & 1           & 7           & 0           & 0           & 0           & 2          & \multicolumn{1}{c|}{0}              & \multicolumn{1}{c|}{0}              & \multicolumn{1}{c|}{0}              & \multicolumn{1}{c|}{0}           & \multicolumn{1}{c|}{2}             & \multicolumn{1}{c|}{11}            & \multicolumn{1}{c|}{8}           & \multicolumn{1}{c|}{9}            & \multicolumn{1}{c|}{1}        & \multicolumn{1}{c|}{\textbf{C'5}}     \\ \hline
\multicolumn{1}{|c|}{\textbf{C6}} & 1           & 14          & 1           & 10          & 0           & 0           & 11          & 0           & 0           & 0           & 0          & \multicolumn{1}{l}{}                & \multicolumn{1}{l}{}                & \multicolumn{1}{l}{}                & \multicolumn{1}{l}{}             & \multicolumn{1}{l}{}               & \multicolumn{1}{l}{}               & \multicolumn{1}{l}{}             & \multicolumn{1}{l}{}              & \multicolumn{1}{l}{}               \\ \cline{1-12}
\end{tabular}
}
\caption{Distribution of students over the clusters of \cref{fig:clusters}.}
\label{table:table}
\end{table}

In order to discuss the results of \cref{fig:clusters}, let us note $(C_i)_{1\leq i \leq 6}$ the identified clusters for the primary school data set, and $(C'_i)_{1\leq i \leq 5}$ for the preparatory classes data set. The order displayed  in \cref{fig:clusters} is from top $(C_1$ and $C'_1)$ to bottom $(C_6$ or $C'_5)$, with the same convention as in \cref{table:table}.

About \cref{clusters-prim}, focusing only on the dense parts and excluding the period from 11:30am to 2pm, we can see that the different clusters can be grouped into three different subsets: $\{C1,C2\}$, $\{C3,C4\}$, and $\{C5,C6\}$ depending on their moments of recess. Indeed, students in the $\{C3,C4\}$ clusters, for example, have a morning recess which is delayed by about an hour compared to the other clusters. As for $\{C1,C2\}$ and $\{C5,C6\}$, they do have morning recess which takes place at the same moment (there is a large overlap even if the duration are not the same) but delayed recess in the afternoons, which made it possible to distinguish them. 
\newline Finally, the two clusters within each of the three subsets are distinguished by the presence or absence of the pupils during the lunch break. In fact, some pupils take their lunches in the school canteen (and we can see that this is a time of important activity because it is dominated by the dense part) while others seem to leave the school compound (passive state during lunch break) to go for lunch elsewhere. 
\newline These observations are confirmed by the data of \cref{table:table}. Except for teachers and students of the $1A$ class, students in the other classes are mainly distributed over two of the identified clusters, depending on their recess time and whether or not they are present inside the school during the lunch break. Each of the clusters, in turn, contains mainly students from classes that had their breaks (morning and afternoon) at the same times. Thus, $\{C1,C2\}$ contains mainly pupils from classes $5A$ and $5B$, $\{C3,C4\}$ those from classes $2A$, $3A$, $3B$ and $4B$. As for the $\{C5,C6\}$ groups, they contain mainly students from classes $1B$, $2B$ and $4A$. The teachers and students of the $1A$ classes who do not stay at school during the lunch break are divided into different groups. 
\newline At this level, it is important to emphasise two points: the first is that the result of clustering obtained on the primary school dataset is mainly dictated by the moments of activation of the dense part on the one hand, and the passivity of the pupils on the other. The reason is that the sparse part is not very present for this dataset and is not concentrated on specific moments of the day. 
\newline The second important point lies in the fact that finding pupils from the same classes within the same clusters is again only due to the nature of the dataset being studied.

About \cref{clusters-lycee}, the  identified clusters  are separated according to properties that can be listed as follows: $C'_1$ is composed of all the nodes that show a high activity (dense part) during the morning break, and a high passivity during the rest of the day. $C'_2$ on the other hand is composed of students characterised by a passivity that extends over most of the day (except for two students who have a long presence in the dense part during the afternoon). $C'_3$ is composed of students who show little activity during the morning period, without being completely passive. In fact, the morning period of this cluster is characterised by students mainly belonging to the sparse part, whereas the afternoon is globally dominated by passive states. It can also be noted that these pupils do not show a high level of activity during the recess (this is artificially the case for class $2BIO3$ whose morning recess is shifted a little later than other classes). It is moreover this last criterion that differentiates the pupils in $C'_3$ from those in $C'_5$ who, in addition to having a high membership to the sparse part during the first part of the day and a strong passivity during the second, are highly present in the dense part during the morning break. Finally, the $C'_4$ cluster is characterised by students with a high membership to the dense part during the second hour of the lunch break. The profiles of these students over the rest of the day are varied.  

Unlike clusters of the primary school where each cluster is heavily dominated by some classes, it is not the case for clusters of high-school data. In the same way, the pupils of each class are more homogeneously distributed over the clusters (except for $2BIO1$ which is almost entirely in the $C'_2$ cluster due to the time lag of its recess). This is on the one hand due to the fact that the pause took place at the same time for the majority of the classes, which pushes the k-means algorithm to carry out its separation by also taking the contribution of the signals $R_S^{(u)}$ and $R_P^{(u)}$. These last signals happen to be characteristic of class periods when activity is low and varies little from one class to another. 

We also note that the size of the clusters varies more on this data set than on the previous one (from 18 students for the smallest to 138 for the largest, against 26 and 54 for the primary school data set) which shows that our measure of similarity defined by \cref{eq:sim} does not induce a bias with respect to the size of the clusters identified by k-means. The two largest clusters $C'_2$ and $C'_3$ together cover two-thirds of the students. So, the majority of the students in the preparatory classes are either in the cluster that shows the greatest passivity or in the one characterised by lack of density during the morning recess.

\begin{figure}[!h]
  \begin{subfigure}[b]{0.49\textwidth}
    \resizebox{1.\textwidth}{.31\textheight}{\includegraphics{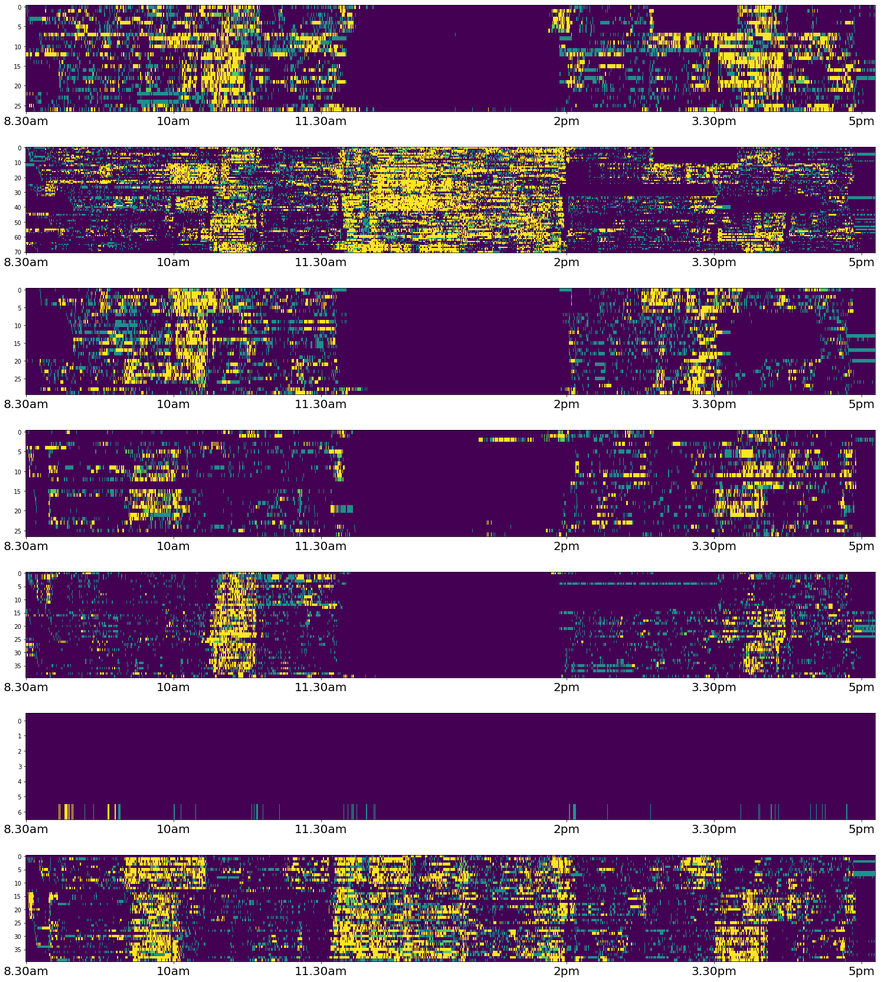}}
    \caption{Primary school data}
    \label{clusters-prim_deg}
  \end{subfigure}
  \begin{subfigure}[b]{0.49\textwidth}
    \resizebox{1.\textwidth}{.31\textheight}{\includegraphics{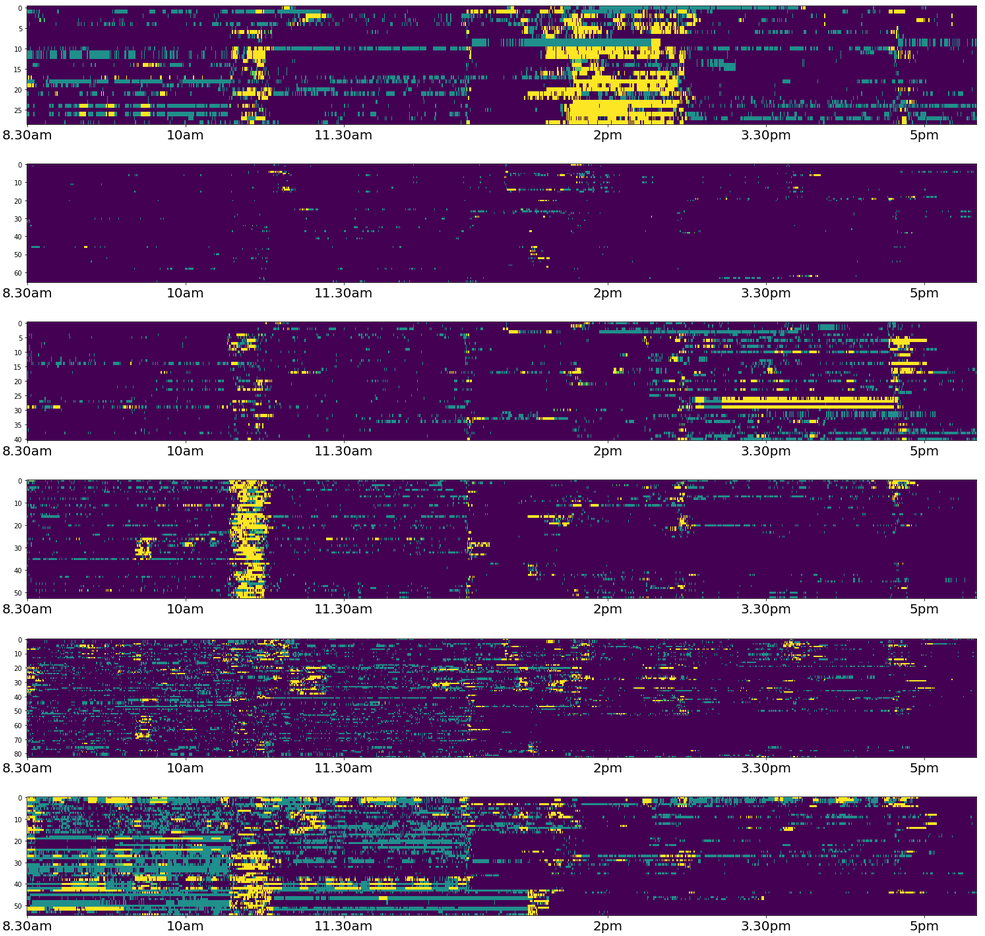}}
    \caption{Highschool data }
    \label{clusters-lycee_deg}
  \end{subfigure} 
 \begin{subfigure}[b]{0.49\textwidth}
    \resizebox{1.\textwidth}{.31\textheight}{\includegraphics{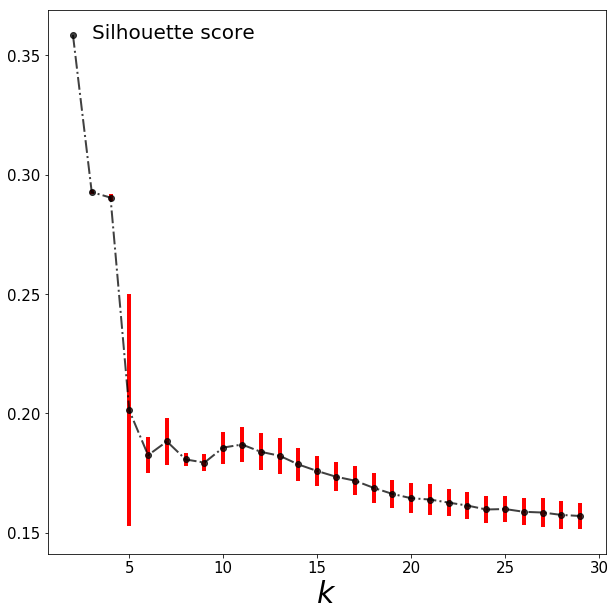}}
    \caption{Primary school data}
    \label{silhouette-prim_deg}
  \end{subfigure}
 \begin{subfigure}[b]{0.49\textwidth}
    \resizebox{1.\textwidth}{.31\textheight}{\includegraphics{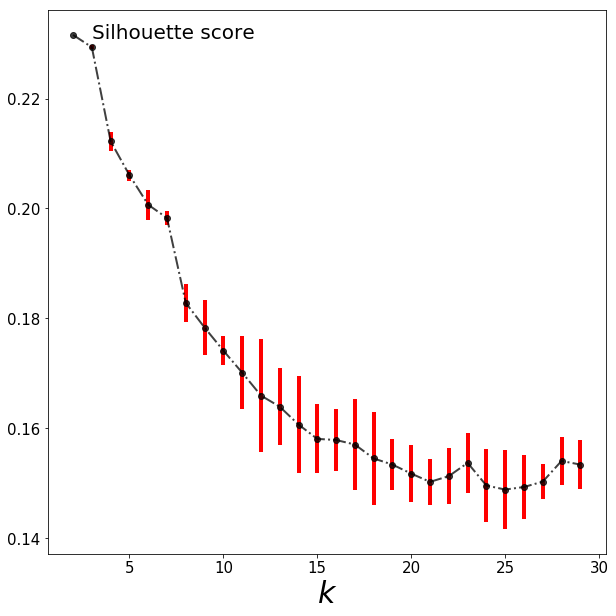}}
    \caption{highschool school data}
    \label{silhouette-lycee_deg}
  \end{subfigure}

\caption{The temporal profiles of students grouped according to clusters obtained using a similarity measure based on the temporal degree of nodes.}
\label{fig:appendix}
\end{figure}

Finally, we would like to note that a classification based only on the degree of a node and not on the triplet $(D,S,P)$ would give worse results in classifying individuals based on temporal and volume similarities of their interactions. To prove that, for each node $u$ we have considered the vector $(d_t(u))_{t\in T}$ of its instantaneous degrees and applied the same method. The description of the results has been pushed in appendix.  
We observe that the obtained partition gives quite different results from those of the $(D,S,P)$-profiles.
Previous results for primary school pupils show that there are at least three types of profiles among the students that have lunch at school depending on their recess time. Results of a classification based on the degree (\cref{clusters-prim_deg}) is unable to separate these profiles.
About the preparatory classes data set, we can see that the partition based on the degree displays 3 clusters similar to those displayed on \cref{fig:clusters}: the first, the second and the fourth clusters starting from the top of \cref{fig:appendix}. But there are however some major differences, notably by observing the last cluster of the \cref{fig:appendix} of the appendix which is composed partly by students who are present in the dense part during the recess, and partly by students who are not, while these students are well separated in a classification on $(D,S,P)$-profiles.

\section{Conclusion}

We described a framework that extend to stream graphs iterative weighted-rich-clubs characterisation for static networks. We used this framework with weights on edges and nodes that are proposed in \cite{dje20}  and dependent on the topological configurations of the node and edge neighbourhoods at each instant, but other values of weights could be chosen. 
Within a temporal context, the general principle is that we no longer consider the membership of a node to one of the weighted-rich-clubs for the whole time period, but each node is associated with a temporal profile which is the concatenation of the successive memberships of the node to the weighted-rich-clubs that appear, disappear and change all along the period. 
In our case, a node can have 3 states at instant $t$ that are provided using algorithm $ItRich$ described on \cite{dje20}: $D$ if it belongs to one of the weighted-rich-clubs existing at this time, $S$ if it does not, and $P$ if it has no interactions at all at time $t$. 
To avoid large discontinuities that may appear between two successive network configurations if the network has few connections, successive states of the nodes are calculated from an average of the weights over small time-windows whose width is a parameter to be chosen. 
The choice of this parameter is driven by the precision of the analysis we want to perform. Please note that in our analysis, the whole time period is just split into such small time-windows but another choice would be to have rolling-windows centred on each current instant.

The approach has been tested on real world data sets, corresponding to temporal networks that record the activity of students in a primary school and preparatory classes during one day. 
It provided a partition of the sets of nodes composing the two temporal networks, with each cluster containing pupils with similar profiles of activity.  
The results are compatible with the ground truth, in particular by separating individuals according to their lunch or recess time. It should be noted, however, our method is not designed to detect temporal communities (in the usual sens of subsets of densely connected nodes) as it is not focused on the internal structure of the clusters.
The usefulness of this clustering approach lies in the fact that it gives the possibility of establishing a reduced list of typical temporal profiles of activity, which is, for example, much more meaningful than a list of typical volumes of activity. 
The method can be refined by not simply considering a division into three states Dense / Sparse / Passive but by taking into account the different weighted rich-clubs that are identified. As the number of weighted-rich-clubs can be different from one instant to another, this would require a method allowing to integrate these different classifications.

\section{Appendix}
In this appendix, we display the result of the partition obtained using a cosine similarity measure computed on the vectors consisting of the temporal degree $d_t$  of each node. We also give the silhouette score for each value of the number $k$  of clusters.
 In the case of primary school data set, we followed  the same criterion as that adopted previously and chose $k=7$ which is a local optimum of the silhouette curve. 
For the preparatory classes data set, we choose to calculate a partition with $k=6$ clusters, which is the same number used to calculate the results displayed on \cref{clusters-lycee}. Indeed, the first local maximum for this data set is obtained for a value of $k=23$ which makes comparison difficult. 


\newpage 
\bibliographystyle{unsrt}
\bibliography{biblio.bib}
\end{document}